\documentclass{PoS}

\title{Equation of state near the first order phase transition point of SU(3) gauge theory using gradient flow}

\ShortTitle{Equation of state near 1$^{st}$ order transition of SU(3) gauge theory using gradient flow}

\author{\speaker{Mizuki Shirogane}%
\\
Graduate School of Science and Technology, Niigata University, Niigata 950-2181, Japan
E-mail: \email{shirogane@muse.sc.niigata-u.ac.jp}
}
\author{Shinji Ejiri \\
Department of Physics, Niigata University, Niigata 950-2181, Japan
}
\author{Ryo Iwami\\
Track Maintenance of Shinkansen, Rail Maintenance 1st Department, East Japan Railway Company Niigata Branch, Niigata, Niigata 950-0086, Japan
}
\author{Kazuyuki Kanaya
\\
Tomonaga Center for the History of the Universe, University of Tsukuba, Tsukuba 305-8571, Japan
}
\author{Masakiyo Kitazawa\\
Department of Physics, Osaka University, Osaka 560-0043, Japan\\
J-PARC Branch, KEK Theory Center, Institute of Particle and Nuclear Studies,
KEK, 203-1, Shirakata, Tokai, Ibaraki, 319-1106, Japan
}
\author{Hiroshi Suzuki\\
Department of Physics, Kyushu University, 744 Motooka, Fukuoka 819-0395, Japan
}
\author{Yusuke Taniguchi\\
Center for Computational Sciences, University of Tsukuba, Tsukuba, Ibaraki 305-8571, Japan
}
\author{Takashi Umeda\\
Graduate School of Education, Hiroshima University, Higashihiroshima, Hiroshima 739-8524, Japan
}

\abstract{
We study energy gap (latent heat) between the hot and cold phases at the first order phase transition point of the SU(3) gauge theory. Performing simulations on lattices with various spatial volumes and lattice spacings, we calculate the energy gap by a method using the Yang-Mills gradient flow and compare it with that by the conventional derivative method.
}

\FullConference{The 36th Annual International Symposium on Lattice Field Theory - LATTICE2018\\
		22-28 July, 2018\\
		Michigan State University, East Lansing, Michigan, USA.}

\begin{document}

\section{Introduction}\label{intro}

We discuss methods to calculate thermodynamic quantities near first order phase transitions.
First order phase transitions are expected in interesting systems including 
the high density region of QCD and the many-flavor QCD aiming at construction of a walking technicolor model.
The SU(3) gauge theory, \textit{i.e.}, the quenched approximation of QCD, at finite temperature is known to have a first order deconfining phase transition and is a good testing ground for developing techniques to investigate thermodynamic quantities around the phase transition.
We study thermodynamic properties near the first order phase transition of the SU(3) gauge theory. 

At a first order phase transition point, two phases coexist. 
To keep a balance between them, their pressures must be the same in the two phases.
Therefore, one can check the reliability of the computational method by measuring the difference of the pressures in two phases.
On the other hand, the energy density is different in each phase.
The difference is the latent heat, which is one of the most important physical quantities characterizing the first order phase transition.
In Ref.~\cite{ejiri98, shirogane16}, we have studied the first order transition of the SU(3) gauge theory using the derivative method \cite{karsch}.
Showing that the pressure gap is absent when we adopt non-perturbative Karsch coefficients, 
we have computed the latent heat, which would be a good reference in developing new methods. 

In this study, we examine a new technique to calculate thermodynamic quantities using the gradient flow, 
proposed by Ref.~\cite{Suzuki:2013gza}. 
In the next section, we introduce the gradient flow method.
Then, our simulations at the first order transition of SU(3) gauge theory is explained in Sec.~\ref{simulation}. We show the results of the latent heat and pressure gap in Sec.~\ref{results} and compare the results with those by the derivative method.
The conclusions are given in Sec.~\ref{conclusion}

\section{Gradient flow method}\label{method}

The gradient flow is an imaginary evolution of the system into a fictitious ``time'' $t$ \cite{Narayanan:2006rf, Luscher:2010iy}. 
We construct a flowed field $B_{\mu}$ solving a kind of diffusion equation and $B_{\mu}$ can be regarded as a smeared field of the original gauge field $A_{\mu}$ over a physical range $\sqrt{8t}$.
It was shown that operators in terms of $B_{\mu}$ have no ultraviolet divergences nor short-distance singularities at finite $t$.
Therefore, the gradient flow defines a physical renormalization scheme, which can be calculated directly on the lattice.
Suzuki proposed a method to calculate the energy- momentum tensor (EMT) making use of the finiteness of the gradient flow \cite{Suzuki:2013gza}. 
In \cite{Asakawa:2013laa, Kitazawa:2016dsl}, the method has been shown to work for the calculation of 
energy density and pressure in quenched QCD. 
Application of the method to full QCD is formulated in Ref.~\cite{Makino:2014taa} and has been successfully performed in Refs.~\cite{WHOT2017b, WHOT2017}.

We calculate the latent heat and pressure gap by the gradient flow method at the first order phase transition point in quenched QCD.
Consider the following gauge-invariant local operators,
\begin{eqnarray}
U_{\mu\nu}(t,x)\equiv G_{\mu\rho}(t,x)G_{\nu\rho}(t,x)
-\frac{1}{4}\delta_{\mu\nu}G_{\rho\sigma}(t,x)G_{\rho\sigma}(t,x),
\ \ \ E(t,x)\equiv\frac{1}{4}G_{\mu\nu}(t,x)G_{\mu\nu}(t,x),
\end{eqnarray}
where $G_{\mu \nu}(t,x)$ is the field strength of flowed gauge field $B_{\mu}$ at the flow time $t$
and the square of $G_{\mu \nu}(t,x)$ is defined by the clover-shaped operator.
With these dimension-$4$ operators, the correctly renormalized energy-momentum tensor $T_{\mu \nu}^R$ is given by \cite{Suzuki:2013gza}
\begin{eqnarray}
T_{\mu\nu}^R(x)
=\lim_{t\to0}\left\{\frac{1}{\alpha_U(t)}U_{\mu\nu}(t,x)
   +\frac{\delta_{\mu\nu}}{4\alpha_E(t)}
   \left[E(t,x)-\left\langle E(t,x)\right\rangle_0 \right]\right\},
\label{eq:(4)}
\end{eqnarray}
where 
$
\alpha_U(t) = \bar{g}^2
\left[1+2b_0 \bar{s}_1 \bar{g}^2+O(\bar{g}^4)\right]
$
and
$
\alpha_E(t) = \frac{1}{2b_0}\left[1+2b_0 \bar{s}_2
\bar{g}^2+O(\bar{g}^4)\right]
$
are determined by the perturbation theory. 
Here, $\bar{g} = \bar{g}(1/\sqrt{8t})$ denotes the running gauge coupling in the
$\overline{\rm MS}$ scheme at the momentum scale $q=1/\sqrt{8t}$, and
$\bar{s}_1=\frac{7}{22}+\frac{1}{2}\gamma_E-\ln2\simeq -0.08635752993$,
$\bar{s}_2=\frac{21}{44}-\frac{b_1}{2b_0^2}=\frac{27}{484}\simeq0.05578512397$,
$b_0=\frac{1}{(4\pi)^2}\frac{11}{3}N_c$,
$b_1=\frac{1}{(4\pi)^4}\frac{34}{3}N_c^2$,
and~$N_c=3$. 
The energy density and the pressure are obtained from the diagonal elements of EMT, 
\begin{eqnarray}
\epsilon = -\left\langle T_{00}^R(x)\right\rangle, 
\hspace{5mm}
p = \frac{1}{3} \sum_{i=1,2,3}\langle T_{ii}^R(x)\rangle .
\end{eqnarray}
Separating configurations into the hot and cold phases, we calculate the latent heat 
$\Delta \epsilon/T^4 \equiv \epsilon^{\rm (hot)}/T^4 - \epsilon^{\rm (cold)}/T^4$ 
and the pressure gap 
$\Delta p/T^4 \equiv p^{\rm (hot)}/T^4 - p^{\rm (cold)}/T^4$,
where
$\epsilon^{\rm (hot/cold)}$ and $p^{\rm (hot/cold)}$ indicate $\epsilon$ and $p$ 
in the hot or cold phase, respectively.

In the gradient flow method of Ref.~\cite{Suzuki:2013gza}, we first take the continuum limit $a\to0$ at each flow time $t$ to remove errors from the lattice regularization, and then take the extrapolation to $t = 0$ to remove unwanted contamination of higher dimension operators.
In Ref.~\cite{WHOT2017b}, an alternative procedure is proposed to first take $t\to0$ at finite $a$ by identifying a range of $t$ in which contribution of terms singular around $t\approx0$ is negligible, and then take the $a\to0$ extrapolation.
The latter procedure is attractive when the $a\to0$ extrapolation is a major source of errors. 
If the removal of singular contributions is successful, the final results should be insensitive to the order of $a\to0$ and $t\to0$ extrapolations. 
We study the effect of the order of these extrapolations.

Another issue in the study of thermodynamic properties is the possible dependence on the physical volume of the system.
We have to keep the physical volume fixed in the $a\to0$ and $t\to0$ extrapolations to identify finite volume effects.
In our study, $T$ is adjusted to the transition temperature $T_c$ whose physical value is a constant.
In this case, the spatial volume $V=(N_s a)^3$ is fixed in physical units when the aspect ratio $N_s/N_t$ is fixed, 
because $V = N_s^3/(N_t T_c)^3$ with $a = 1/(N_t T_c)$ .
We study the cases of $N_s/N_t=6$ and 8 to study the finite volume effect.

\section{Numerical Simulations}\label{simulation}

We perform simulations of the SU(3) gauge theory with the standard Wilson action at several $\beta$'s around the deconfining transition point. 
We have studied the lattices with temporal lattice size of $N_t=8$, 12 and $16$ with several different spatial lattice size $N_s$ for each $N_t$.
Some of the simulation data are obtained in Ref.~\cite{shirogane16}.
Though we have tested $N_t=6$ lattices too, we do not use them in this study because the $t\to0$ extrapolation was found to be ambiguous for $N_t=6$.
Our simulation parameters are summarized in Table~\ref{tab1}.
The configurations are generated by a pseudo heat bath algorithm followed by 5 over-relaxation sweeps.
Performing the gradient flow, smeared observables are measured every 20 or 50 iterations.
Data are taken at 3 to 7 $\beta$ values for each $(N_s, N_t)$,
and are combined using the multipoint reweighting method \cite{iwami15}.

To evaluate $\Delta \epsilon$ and $\Delta p$, we need to separate the configurations at the first order phase transition point into the hot and cold phases.
We use the method adopted in Ref.~\cite{shirogane16}.
As shown in Ref.~\cite{shirogane16}, 
there are two peaks corresponding to the hot and cold phases in the 2-dimensional histogram of Polyakov loop and paquette at the transition point. 
The peaks are well separated in the Polyakov loop direction, while they are overlapping in the plaquette direction.
We thus classify configurations into the hot and cold phases with the value of the Polyakov loop.
Because configurations in which two phases coexist are found to be rare on our lattices, we disregard the effects of the mixed phase. 

\begin{table}[thb]
  \small
  \centering
  \caption{Simulation parameters: ``\#$\beta$'', ``$\beta_{\rm min}$'' and ``$\beta_{\rm max}$'' are the number of simulation points, the smallest and largest values of $\beta$, respectively. 
The number of Monte Carlo steps are given in the column ``traj'', and the interval of the measurements in the column ``int''.
The results of $t\to0$ extrapolation for $\Delta(\epsilon +p)/T^4$ and $\Delta(\epsilon -3p)/T^4$ on each lattice are shown in the last two column,
while, in the last two lines, the results of $t\to0$ extrapolation after the continuum extrapolation are shown.
}
  \label{tab1}
\vspace{1.5mm}
  \begin{tabular}{rrrrrrrrr}
\hline
$N_s$& $N_t$& \#$\beta$& $\beta_{\rm min}$& $\beta_{\rm max}$& traj& int &$\Delta(\epsilon +p)/T^4$ &$\Delta(\epsilon -3p)/T^4$ \\
\hline
 48&  8& 6& 6.056 & 6.067& 1220000& 20& 1.048(38)& 1.255(46) \\
 64&  8& 5& 6.0585& 6.065& 4535000& 20& 1.035(08)& 1.159(10) \\
 48& 12& 3& 6.333 & 6.337& 4750000& 50& 1.395(20)& 1.440(22) \\
 64& 12& 5& 6.332 & 6.339& 4750000& 50& 1.302(43)& 1.323(46) \\
 96& 12& 7& 6.33  & 6.339& 3747500& 50& 1.035(30)& 1.080(33) \\
 96& 16& 3& 6.543 & 6.547&  720000& 50& 1.198(20)& 1.201(28) \\
128& 16& 3& 6.543 & 6.547&  341000& 50& 0.970(52)& 0.936(46) \\
\hline
 \multicolumn{7}{c}{$N_s / N_t = 8$, continuum limit} & 1.103(47) & 0.983(57) \\
 \multicolumn{7}{c}{$N_s / N_t = 6$, continuum limit} & 1.275(67) & 1.202(59) \\
\hline
  \end{tabular}
\end{table}

\begin{figure}[tbh]
\centering
\includegraphics[width=7cm,clip]{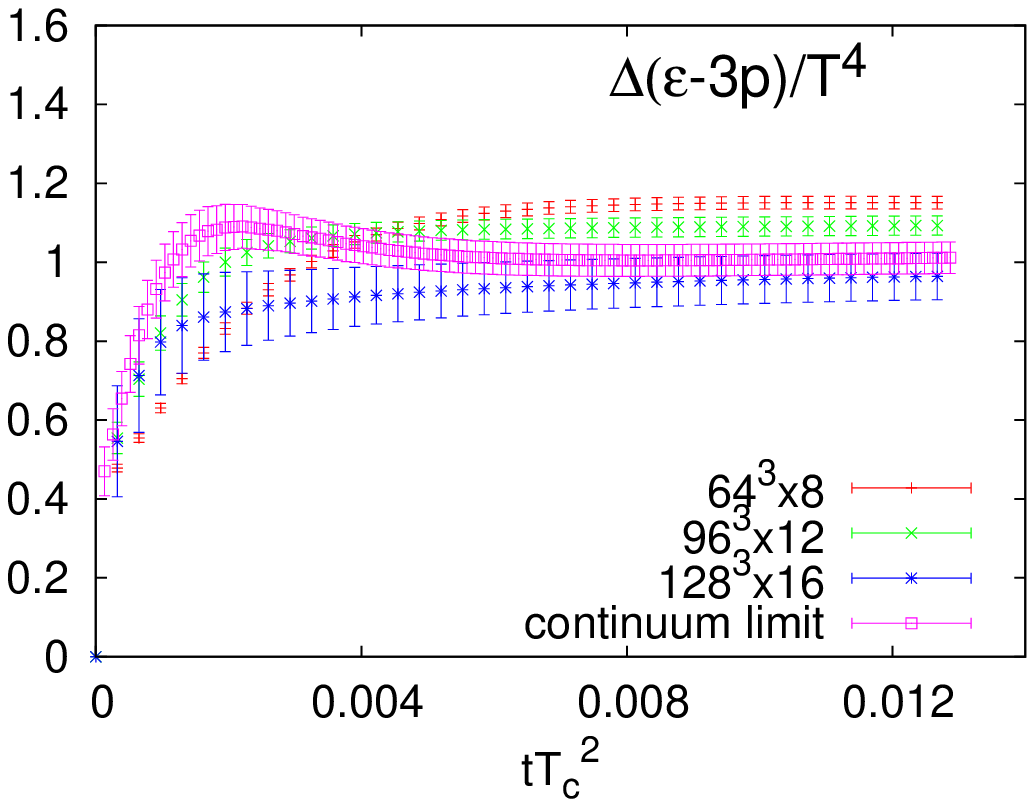}
\hspace{1mm}
\includegraphics[width=7cm,clip]{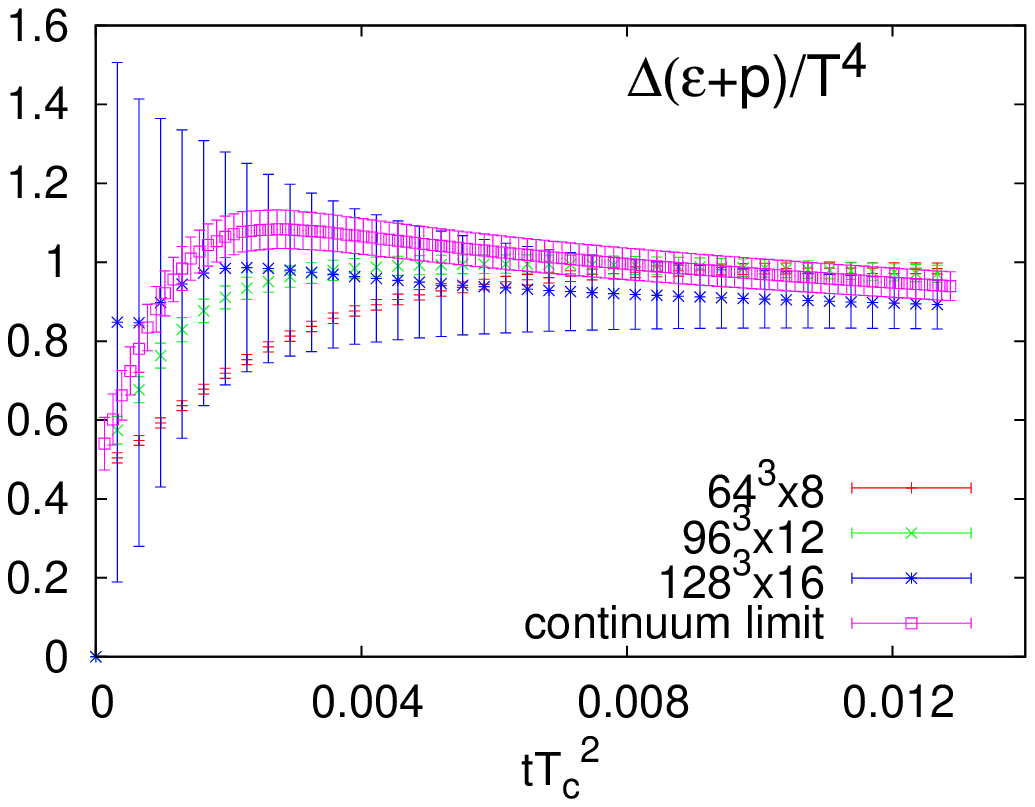}
\vspace{-3mm}
\caption{$\Delta (\epsilon -3p)/T^4$ (left) and $\Delta (\epsilon +p)/T^4$ (right) measured on $N_s/N_t=8$ lattices.
}
\label{fig1}
\end{figure}

\begin{figure}[tbh]
\centering
\includegraphics[width=7cm,clip]{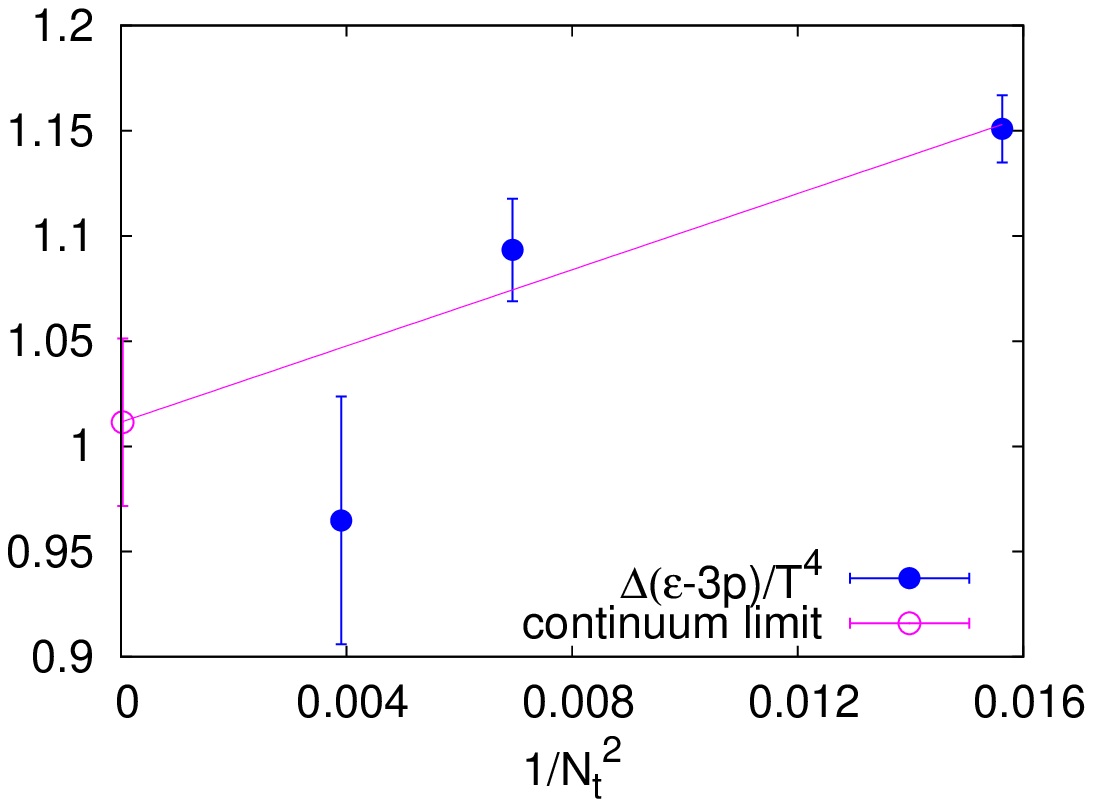}
\hspace{1mm}
\includegraphics[width=7cm,clip]{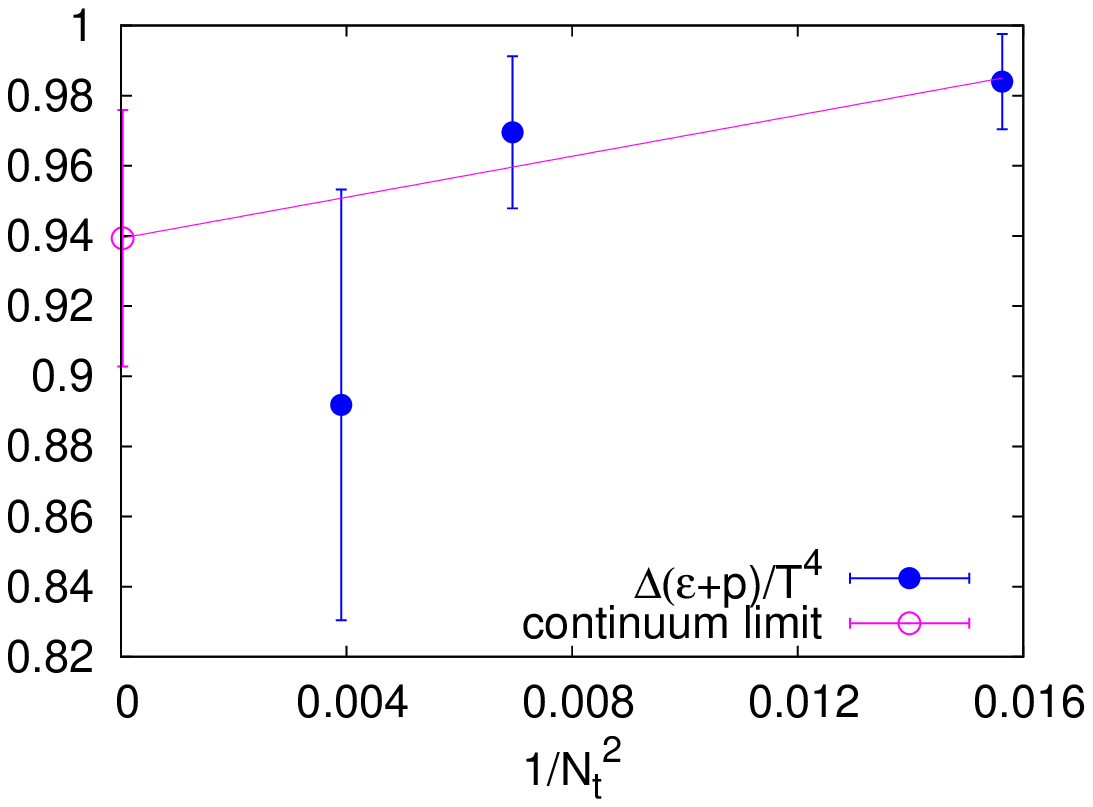}
\vspace{-3mm}
\caption{
$\Delta (\epsilon -3p)/T^4$ (left) and $\Delta (\epsilon +p)/T^4$ (right) as functions of $1/N_t^2 = (T_c a)^2$ for $t T_c^2 = 0.013$ and $N_s/N_t=8$.
}
\label{fig2}
\end{figure}

\begin{figure}[tbh]
\centering
\includegraphics[width=7cm,clip]{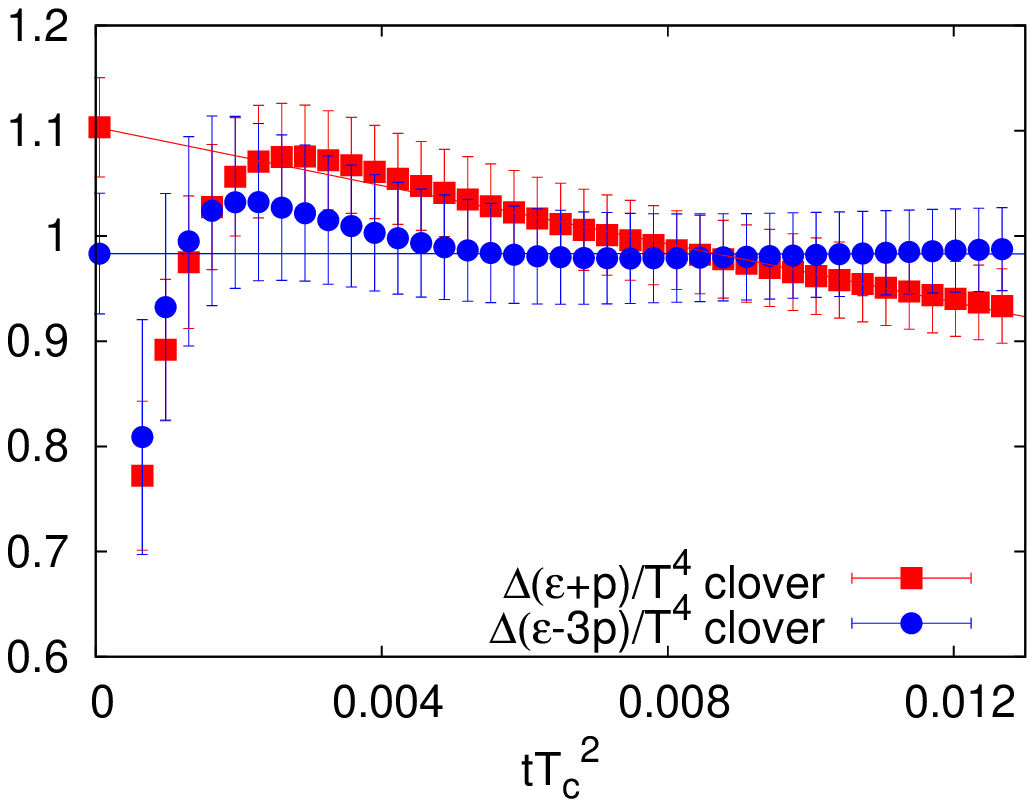}
\hspace{1mm}
\includegraphics[width=7cm,clip]{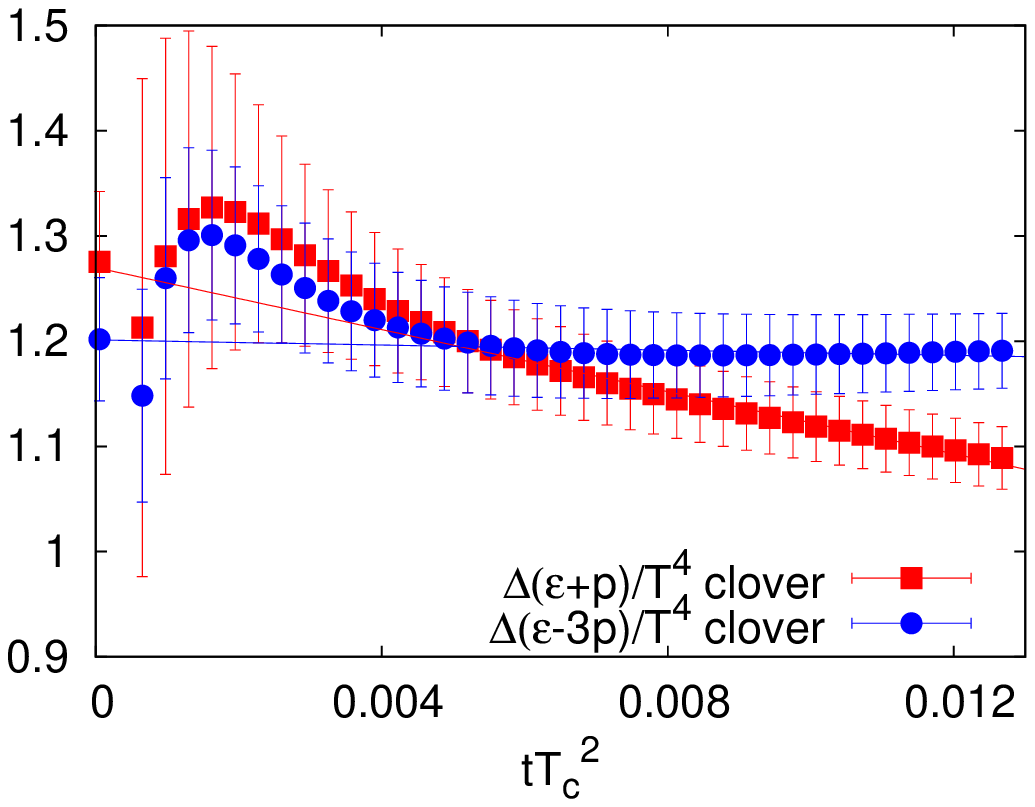}
\vspace{-3mm}
\caption{$\Delta (\epsilon+p)/T^4$ and $\Delta (\epsilon-3p)/T^4$ 
in the continuum limit measured on 
lattices with $N_s/N_t=8$ (left) and $N_s/N_t=6$ (right).
The symbols on the vertical axis indicate the results of $t\to0$ extrapolations.
}
\label{fig3}
\end{figure}

\section{Results of the equation of state}\label{results}

In Fig.~\ref{fig1}, we show the results of $\Delta(\epsilon -3 p)/T^4$ (left) and $\Delta(\epsilon + p)/T^4$ (right)
as functions of $t T^2_c =t/(aN_t)^2$ on lattices with $N_s/N_t=8$.
The red, green and blue symbols are the results of $N_t=8$, $12$ and $16$, respectively.
As $t$ is increased, the lattice discretization error decreases because the smearing length becomes larger than the lattice spacing.
In fact, the plots in Fig.~\ref{fig1} show that the difference among the results with different lattice spacing becomes smaller as the flow time $t$ increases, 
and the results of $N_t=8$, 12 and 16 are consistent in the range of $t T_c^2 \geq 0.006$ for $\Delta (\epsilon +p) /T^4$.

We first adopt the original procedure of taking $a\to0$ first and then $t\to0$.
Examples of the $a\to0$ extrapolation are shown in Fig.~\ref{fig2}.
We fit the data with a linear function of $a^2 \propto 1/N_t^2$. 
The magenta symbols in Fig.~\ref{fig1} are the results in the continuum limit.
We repeat the analyses also for $N_s/N_t=6$ using the data of $N_t=8$ and 16 lattices.

In Fig.~\ref{fig3}, we show the results of $\Delta(\epsilon + p)/T^4$ and  $\Delta(\epsilon -3 p)/T^4$ in the continuum limit as functions of $tT_c^2$ on the lattices with $N_s / N_t =8$ (left) and $N_s / N_t =6$ (right). 
Omitting the data at small $t$ where the lattice artifact is still large, 
we fit the data at $0.004 < tT_c^2 < 0.013$ by a liner function to extract the $t=0$ limit.
We find that $\Delta (\epsilon +p) /T^4$ and $\Delta (\epsilon -3p) /T^4$ at $t = 0$ are consistent with each other within about one sigma, 
suggesting $\Delta p = 0$ on these lattices.

We now study the alternative procedure of taking $t\to0$ first on each lattice and then take $a\to0$. 
We carry out the $t\to0$ extrapolations by a linear fit in $t T_c^2$ adopting the fit range  
$0.008 < t T_c^2 < 0.013$ for $N_s/N_t=8$ and 
$0.01 < t T_c^2 < 0.016$ for $N_s/N_t \leq 6$. 
The results of $\Delta(\epsilon + p)/T^4$ and $\Delta(\epsilon -3 p)/T^4$ at $t=0$ are summarized in Table~\ref{tab1} for each lattice. 
The errors are statistical only.
We find that the differences between $\Delta (\epsilon -3p) /T^4$ and $\Delta (\epsilon +p) /T^4$ are less than about one sigma for $N_t=12$ and $16$ lattices, indicating $\Delta p \approx 0$ on these finite lattices, 
while the differences for $N_t=8$ are larger than the statistical error. 
Note that the systematic errors including those due to the choice of the fit range for the $t\to0$ extrapolation, are not included yet.

\begin{figure}[tbh]
\centering
\includegraphics[width=7cm,clip]{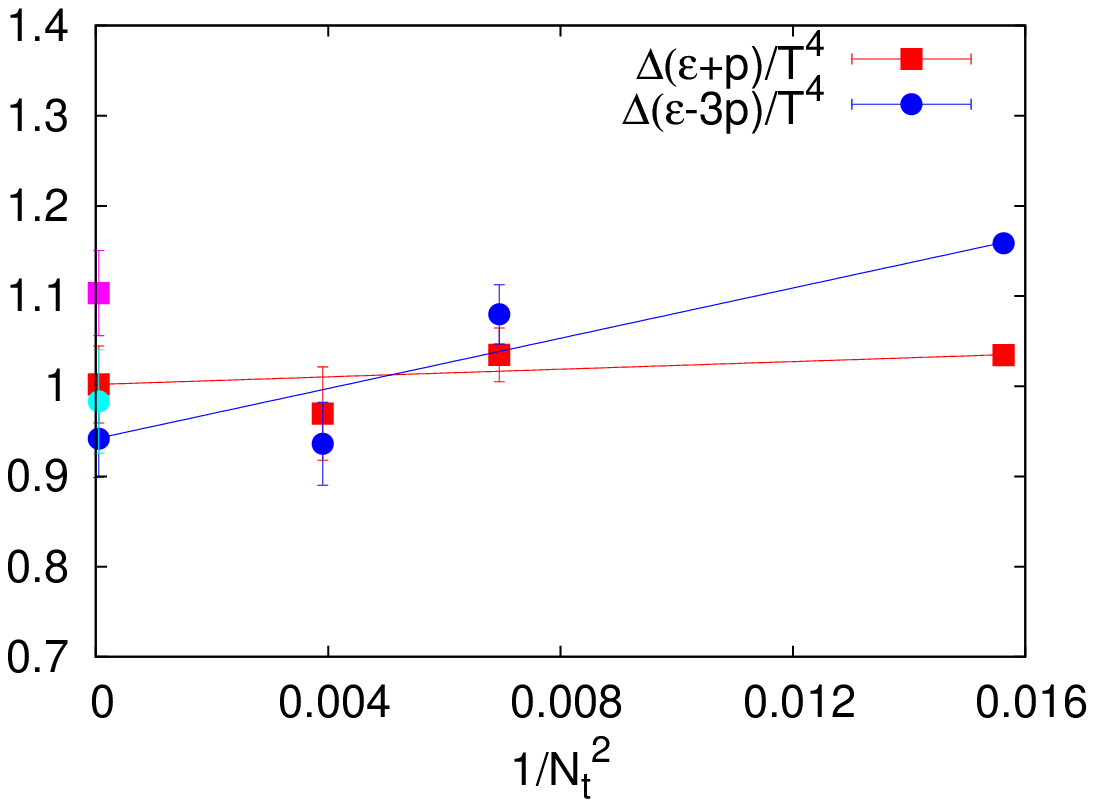}
\hspace{1mm}
\includegraphics[width=7cm,clip]{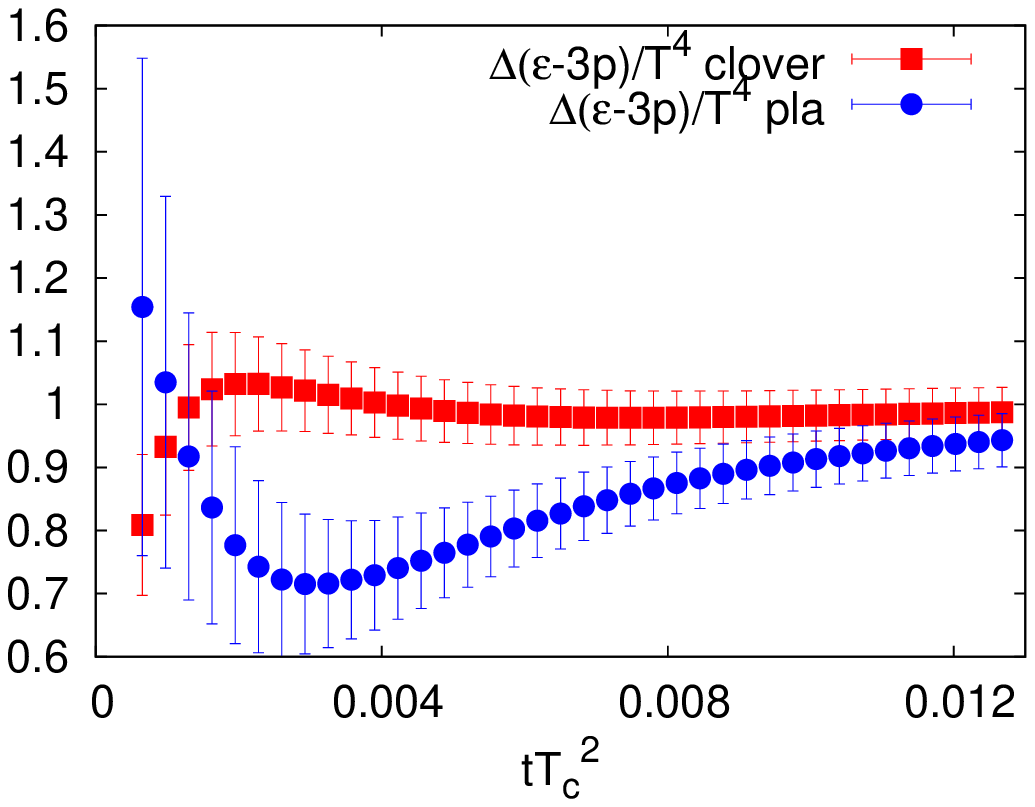}
\vspace{-3mm}
\caption{\textbf{Left}: $\Delta (\epsilon+p)/T^4$ (squre) and $\Delta (\epsilon-3p)/T^4$ (circle) at $t=0$ measured on $N_s/N_t=8$ lattices. 
The horizontal axis is $1/N_t^2 = (T_c a)^2$
The magenta and cyan symbols at $1/N_t^2 = 0$ are the results of the original procedure of first taking $a\to0$ and then $t\to0$.
\textbf{Right}: $\Delta (\epsilon-3p)/T^4$ using clover-shaped operator (red) and plaquette (blue) in the continuum limit for $N_s/N_t=8$.
}
\label{fig4}
\end{figure}

\begin{figure}[tbh]
\centering
\includegraphics[width=7cm,clip]{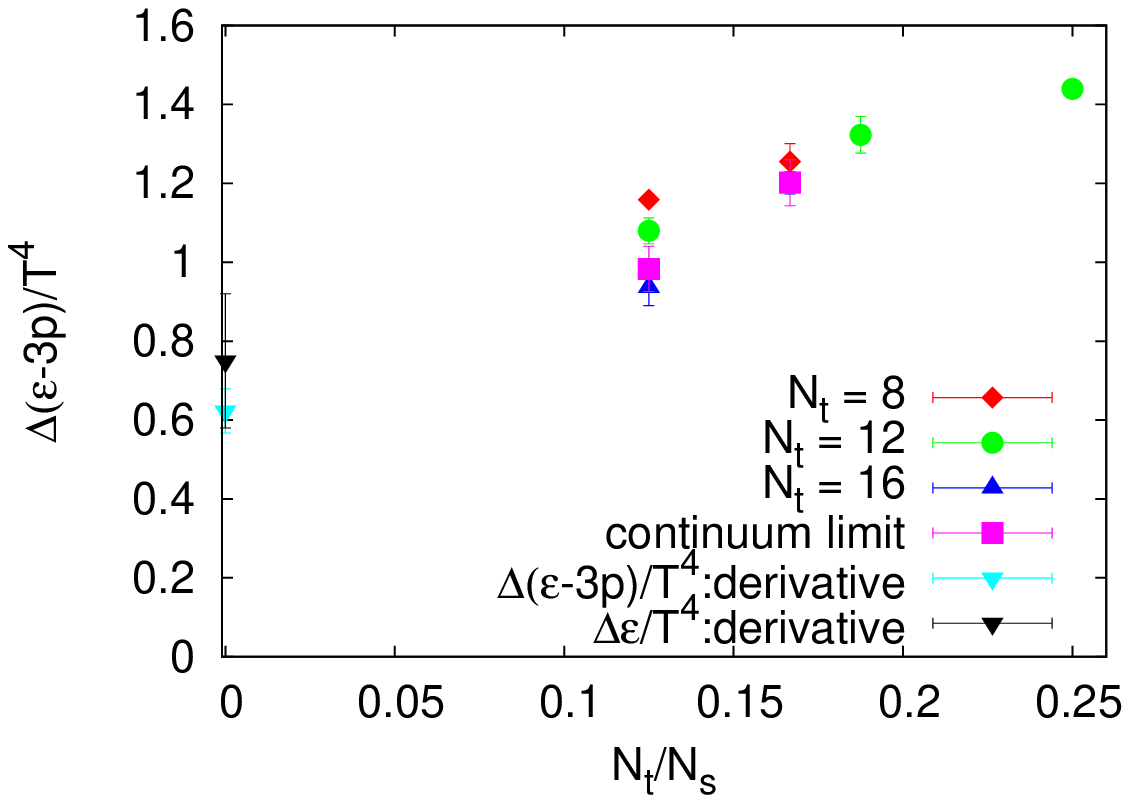}
\hspace{1mm}
\includegraphics[width=7cm,clip]{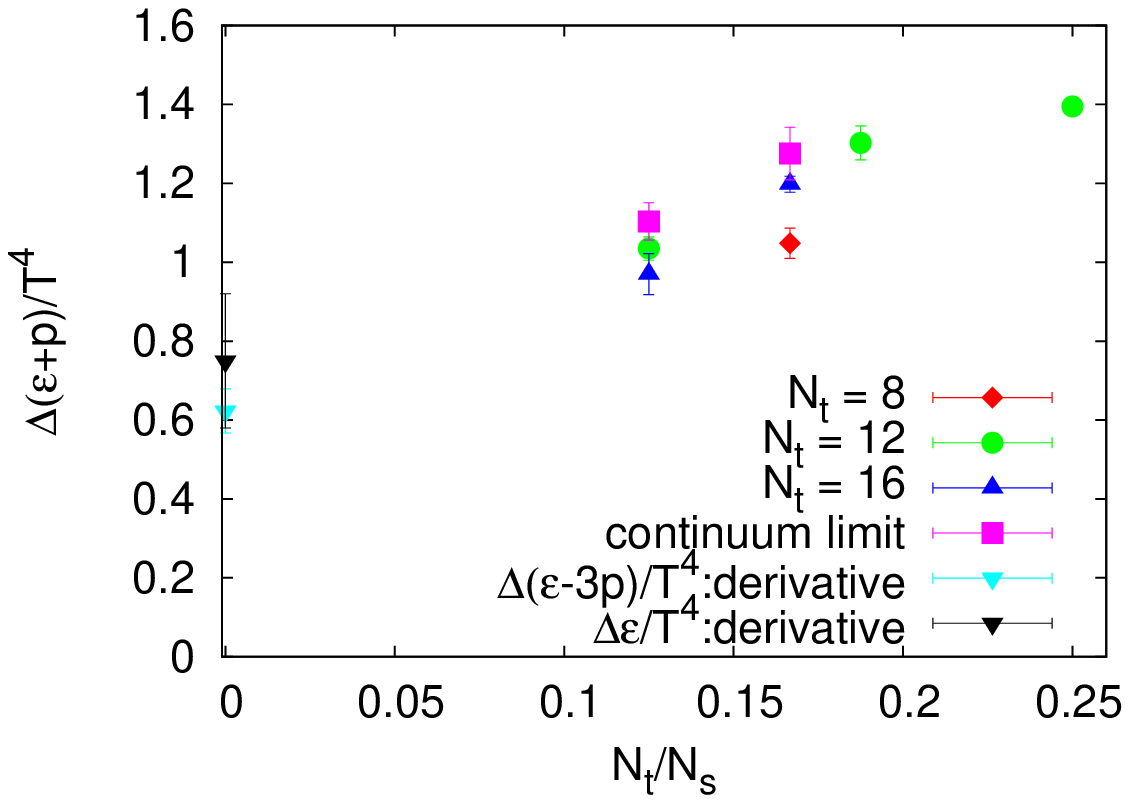}
\vspace{-3mm}
\caption{$\Delta (\epsilon-3p)/T^4$ (left) and $\Delta (\epsilon+p)/T^4$ (right) at $t=0$ as functions of $N_t /N_s =1 /(\sqrt[3]{V} T_c)$.
The black and cyan triangles on the vertical axis are the result of the derivative method in the continuum limit \cite{shirogane16}
}
\label{fig5}
\end{figure}

In the left panel of Fig.~\ref{fig4}, we plot $\Delta (\epsilon -3p) /T^4$ and $\Delta (\epsilon +p) /T^4$ at $t=0$ as functions of $1/N_t^2 = (T_c a)^2$ for $N_s /N_t =8$. 
The symbols on the vertical axis are the result at $t=0$.
The magenta and green symbols on the vertical axis are the results of the original procedure of $a\to0$ followed by $t\to0$.
We see that the results obtained by the two procedures are consistent with each other within the statistical errors.

Here, we comment on the choice of the operator for $G_{\mu \nu}^2$. 
So far, we have discussed the results of $\Delta \epsilon$ and $\Delta p$ computed by the clover-shaped operator for $G_{\mu \nu}^2$. 
However, $G_{\mu \nu}^2$ can also be computed by the plaquette operator.
In the right panel of Fig.~\ref{fig4}, we compare the results of $\Delta(\epsilon -3 p)/T^4$ in the continuum limit obtained by the clover-shaped (red) and plaquette (blue) operators for $N_s/N_t=8$. 
The two results seem to agree with each other at sufficiently large $t$. 

Finally, we study the finite volume effect and compare our results of the gradient flow method with those obtained by the derivative method \cite{shirogane16}.
In Fig.~\ref{fig5}, we plot the results of $\Delta(\epsilon + p)/T^4$ and $\Delta(\epsilon -3 p)/T^4$ 
in the $t\to0$ limit as functions $N_t /N_s = 1 /(\sqrt[3]{V} T_c)$.
The magenta symbols are the results of in the continuum limit for $N_s /N_t =6$ and 8 obtained by the original procedure of $t\to0$ after $a\to0$.
For comparison, we also show the results at $t=0$ on lattices with $N_t =8$, 12 and 16 by red, green and blue symbols, respectively.
The black and cyan triangles on the vertical axis are the results of the derivative method for 
$\Delta \epsilon /T^4$ and $\Delta (\epsilon -3p) /T^4$, respectively \cite{shirogane16}. 
We find that the latent heat by the gradient flow method decreases and approaches the results of the derivative method as the physical volume increases. 
Here, however, it should be kept in mind that, as suggested by Figs.~8 and 9 of Ref.~\cite{shirogane16}, we may have sizable systematic errors in the results of the derivative method due to the infinite volume and continuum extrapolations.

\section{Conclusions}\label{conclusion}

We calculated the latent heat and the pressure gap between the hot and cold phases at the first order phase transition point of the SU(3) gauge theory using the gradient flow method,
performing simulations on lattices with various spatial volumes and lattice spacings.
We confirmed that the pressure gap at the transition point is consistent with zero on fine lattices with $N_t \geq 12$.
We then studied the influence of the order of $t\to0$ and $a\to0$ extrapolations on the latent heat and the pressure gap, 
and found that the results of the original procedure ($a\to0$ followed by $t\to0$) and the alternative procedure ($t\to0$ followed by $a\to0$) are consistent with each other.
We also studied the finite volume effect on the latent heat.
We found that the latent heat by the gradient flow method decreases and approaches the result of the derivative method 
as we increase the physical volume of the system towards the infinite volume limit (thermodynamic limit). 

\vspace{5mm}

This work was in part supported by JSPS KAKENHI (Grant Nos.\ 
JP26287040, JP26400251, JP15K05041, JP16H03982, and JP17K05442),
the Uchida Energy Science Promotion Foundation, 
the Large Scale Simulation Program of High Energy Accelerator
Research Organization (KEK) (No.\ 15/16-25, 16/17-05),
the HPCI System Research project (Project ID: hp170208),
Interdisciplinary Computational Science Program in CCS, University of Tsukuba,
and the large-scale computation program on OCTPUS at the Cybermedia Center, Osaka University.


\end{document}